\documentclass[floatfix,aps,twocolumn,showpacs,preprintnumbers,footinbib,longbibliography,prl,superscriptaddress]{revtex4-1}

\usepackage{mathptmx}

\usepackage[autostyle=true]{csquotes}

\usepackage{mathtools}
\usepackage{textcomp}
\usepackage{gensymb}
\usepackage{graphicx}
\usepackage{siunitx}
\usepackage{ulem}

\usepackage{color}

\usepackage[unicode=true,
 bookmarks=true,bookmarksnumbered=true,bookmarksopen=true,bookmarksopenlevel=2,
 breaklinks=false,pdfborder={0 0 1},backref=false,colorlinks=true]
 {hyperref}

 \hypersetup{linkcolor=blue, citecolor=blue, urlcolor=blue, filecolor=blue, pdfpagelayout=OneColumn,
  pdfnewwindow=true, pdfstartview=XYZ, plainpages=false}

\usepackage[all]{hypcap}

\begin{document}

\def\be{\begin{equation}}
\def\ee{\end{equation}}
\def\ki{\kappa_i}
\def\kf{\kappa_f}
\def\kopt{\kappa_{opt}}
\def\tr{\tau_{relax}}
\def\pN{pN/$\mu$m}

\title{Optimal protocols and universal time-energy bound in Brownian thermodynamics}

\author{Yoseline Rosales-Cabara}
\affiliation{Universit\'e de Strasbourg, CNRS, Institut de Science et d'Ing\'enierie Supramol\'eculaires, UMR 7006, F-67000 Strasbourg, France}
\author{Giovanni Manfredi}
\email[]{giovanni.manfredi@ipcms.unistra.fr}
\affiliation{Universit\'e de Strasbourg, CNRS, Institut de Physique et Chimie des Mat\'eriaux de Strasbourg, UMR 7504, F-67000 Strasbourg, France}
\author{Gabriel Schnoering}
\affiliation{Universit\'e de Strasbourg, CNRS, Institut de Science et d'Ing\'enierie Supramol\'eculaires, UMR 7006, F-67000 Strasbourg, France}
\author{Paul-Antoine Hervieux}
\email[]{hervieux@ipcms.unistra.fr}
\affiliation{Universit\'e de Strasbourg, CNRS, Institut de Physique et Chimie des Mat\'eriaux de Strasbourg, UMR 7504, F-67000 Strasbourg, France}
\author{Laurent Mertz}
\email[]{lm167@nyu.edu}
\affiliation{NYU-ECNU Institute of Mathematical Sciences at NYU Shanghai,, Shanghai, 200062, China}
\author{Cyriaque Genet}
\email[]{genet@unistra.fr}
\affiliation{Universit\'e de Strasbourg, CNRS, Institut de Science et d'Ing\'enierie Supramol\'eculaires, UMR 7006, F-67000 Strasbourg, France}

\date{\today}

\begin{abstract}

We propose an optimization strategy to control the dynamics of a stochastic system transferred from one thermal equilibrium to another and apply it experimentally to a Brownian particle in an optical trap under compression. Based on a variational principle  that treats the transfer duration and the expended work on an equal footing, our strategy leads to a family of protocols that are either optimally cheap for a given duration or optimally fast for a given energetic cost. This approach unveils a universal relation $\Delta t\,\Delta W \ge (\Delta t\,\Delta W)_{\rm opt}$ between the transfer duration and the expended work. We verify experimentally that the lower bound is reached only with the optimized protocols.

\end{abstract}

\maketitle

Controlling the transformation of equilibrium states (and related quantities) is a major concern in stochastic energetics.
Though still in its infancy, this is an important research area with promising applications for nanotechnologies.
Recently, many experimental and theoretical perspectives, both in the classical and quantum regimes \cite{SchmiedlPRL2007,ChenPRL2010,AurellPRL2011,SeifertRPP2012,WeberNature2014}, have demonstrated the possibility to control the evolution of a small system while constraining a set of thermodynamic variables using appropriate protocols. For instance, recent work proposed protocols that can force a nano- or micro-system to evolve from one equilibrium state to another much faster than the relaxation time expected from the energy difference between the two equilibria \cite{MartinezNatPhys2016,ChupeauPRE2018,LeCunuderAPL2016}. From a mathematical viewpoint, this is an interesting optimal control problem, which can be studied using the Pontryagin's principle \cite{MR0166037,PlataPRE2019}.

Accelerated equilibration protocols have direct thermodynamic consequences. A protocol that reduces the transfer duration is necessarily more expensive energetically, so that the requirements of being fast and cheap cannot be satisfied simultaneously \cite{MartinezNatPhys2016}. Earlier proposals have discussed the possibility to minimize the work expended through a transfer whose duration is initially fixed \cite{SchmiedlPRL2007}. But this approach does not treat duration and work on an equal footing: while the former is arbitrarily fixed by the experimentalist, only the latter is minimized. This strategy prevents one from  deriving and exploiting the mutually exclusive relation between a protocol's duration and its energetic cost, which is of paramount importance to design protocols that are optimized from both points of view. The possibility for optimal control turns out to be particularly relevant in the field of stochastic engines, where it is necessarily related to the global figure of merit of the system \cite{BlickleNatPhys2011,RoldanNatPhys2015}.

In order to derive such optimal protocols, we adopt an original approach that treats both the duration of the transfer and the expended work in a completely symmetric way. Our strategy, implemented on an optically trapped Brownian particle,  is based on two novel ingredients. First, each protocol is defined by a path in the phase space $(\kappa, s)$, where $\kappa$ is the stiffness of the optical trap and $s$ is the variance of the position of the particle. Second, we construct a functional $J[\kappa, s]$ that is composed of two terms, corresponding respectively to the total work and total duration, with a Lagrange multiplier $\lambda$ regulating the trade-off between these two quantities. Minimizing the above functional with fixed boundary conditions, leads to the desired optimal protocol $\kappa_{\lambda}(s)$. For instance, $\lambda \gg 1$ yields a protocol that has a low energetic cost but long duration; conversely, $\lambda \ll 1$ leads to a fast protocol that requires a large amount of work.

Remarkably, this approach leads to a universal relation $\Delta t\,\Delta W \ge (\Delta t\,\Delta W)_{\rm opt}$ between the transfer duration $\Delta t$ and the expended work $\Delta W$ (in excess of the free energy difference), where the lower bound depends exclusively on the initial and final states and can be reached only under optimal control conditions. This result unveils a fundamental feature that underpins all optimization procedures in stochastic thermodynamics \cite{SekimotoBook,CilibertoPRX2017}.

Our Brownian particle is a polystyrene microsphere optically trapped in water at room temperature --  see Appendix A for a detailed description of the setup \cite{SchnoeringPRE2015,SchnoeringPRAppl2019,SchnoeringPRL2018}. In this overdamped regime, the conditions are carefully set so that the trapping potential is harmonic. We record the instantaneous motion $x(t)$ of the microsphere along the optical axis of the trap. With the trap stiffness $\kappa$ proportional to the trapping laser intensity $I$, it is possible to define, by modulating $I(t)$, a given protocol $\kappa (t)$ for the transfer from an initial thermal equilibrium at time $t_i$ to a final equilibrium at time $t_f$.

By performing a series of $N$ identical protocols  on the trapped microsphere, we build a statistical ensemble of trajectories that yields a probability density function (PDF) of positions $x$. The dynamics of the system is described by the variance $s(t)$ extracted from the PDF that evolves according to
\be
\gamma\frac{ds}{dt}= -2\kappa(t)s + 2D\gamma,
\label{eq:variance}
\ee
where $\gamma=6\pi R\eta$ is the Stokes drag coefficient, which depends on the radius of the particle $R=500$ nm and the dynamic viscosity of the fluid $\eta\sim 10^{-3} \ {\rm Pa}\,{\rm s}$, and $D=k_B T /\gamma\sim 0.4 \ \mu{\rm m}^2 /{\rm s}$ the Brownian diffusion coefficient fixed by the temperature $T$ of the fluid and the Boltzmann constant $k_B$.

Equation \eqref{eq:variance} fully determines the statistical properties of the system where the initial and final equilibria correspond to the stationary solutions $s_i \kappa_i = s_f \kappa_f = k_B T$ with Gaussian PDF. It is also clear that the PDF will remain Gaussian for all intermediate times between $t_i$ and $t_f$. The cumulative energetics involved during the protocol is directly related to the time evolution of $s(t)$, giving the ensemble-averaged expended work $W(t) =\frac{1}{2}\int_{t_i}^{t}{\rm d}t s(t)\dot{\kappa}(t)$ and dissipated heat $ Q(t)  =-\frac{1}{2}\int_{t_i}^{t}{\rm d}t \dot{s}(t)\kappa(t)$ (following the convention of a positive flow when heat is transferred from the trapped microsphere to the bath) \cite{SekimotoPTP1998,CilibertoPRX2017}.

Our purpose is to control the dynamics of the microsphere so that the transfer between the two equilibria is optimal with respect to both duration and energetics. If one switches instantaneously the trap stiffness from $\kappa_i$ to $\kappa_f>\kappa_i$ (i.e., closing the trap in a step-like way), the typical relaxation time to the new equilibrium is given by $\tau_{\rm relax}=2 \gamma/\kappa_f$. This time will be taken as the reference value with respect to which the reduction in transfer duration is measured.

Our optimization strategy starts with the idea of using the variance $s$ as the independent variable of the problem, instead of the time $t$.
This is possible whenever the function $s(t)$ is monotonic and it enables us to express the control parameter as $\kappa(t)=\hat{\kappa}\left(s(t)\right)$. The advantage of this approach is that we can easily write down, as functionals of $\hat{\kappa}(s)$, both the transfer duration:
\be
\Delta t [\hat{\kappa}(s)] \equiv t_f-t_i = {1\over 2}\,\int_{s_i}^{s_f} \frac{\gamma\,ds }{D\gamma-s\, \hat{\kappa}(s)}\, ,
\label{eq:deltat}
\ee
and the expended work:
\be
 W[\hat{\kappa}(s)]
= -{1\over 2}\int_{s_i}^{s_f}\hat{\kappa}(s) ds + {1\over 2}\left(\kappa_f s_f - \kappa_i s_i\right),
\label{eq:work}
\ee
where the second term on the right-hand side vanishes because the initial and final configurations are thermally equilibrated.

Next, we define the functional to be minimized as twice the sum of $W$ and $\Delta t$:
\be
J[\hat{\kappa}(s)] = \int_{s_i}^{s_f} \frac{\gamma ds }{D\gamma-s\, \hat{\kappa}(s)} - \lambda \int_{s_i}^{s_f}\hat{\kappa}(s) ds ,
\label{eq:J}
\ee
where $\lambda$ is a Lagrange multiplier that regulates the trade-off between transfer duration and work. Within this framework, the optimization strategy can be interpreted as the search for the trajectory in the $(s,\kappa)$-space that minimizes $J[\hat{\kappa}(s)]$ while keeping the extrema fixed at equilibrium, i.e. $s_i \kappa_i = s_f \kappa_f = k_{\rm B}T=D\gamma$.
Once written as $J=\int_{s_i}^{s_f} L[s,\hat{\kappa}(s)]ds$, this functional can be minimized using the standard Euler-Lagrange equation
\be
{d \over{ds}} \frac{\partial L}{\partial \hat{\kappa}'} - \frac{\partial L}{\partial \hat{\kappa}}=0,
\label{eq:EL}
\ee
where $\hat{\kappa}' \equiv d\hat{\kappa}/ds$, yielding the following solution
\be
s\,\hat{\kappa}(s) = D\gamma + \sqrt{\gamma s / \lambda}
\label{eq:kappa-s}
\ee
for a protocol that eventually closes the trap with $\kappa_f > \kappa_i$ \footnote{In order to satisfy $\Delta t>0$, a negative sign should be used on the r.h.s. of Eq. \eqref{eq:kappa-s} when $s_f>s_i$ (opening trap, $\kappa_f < \kappa_i$), whereas a positive sign should be used when $s_f < s_i$ (closing trap, $\kappa_f > \kappa_i$).}.

Equation \eqref{eq:kappa-s} encapsulates the main result obtained so far. For instance, the quasi-static solution ($\dot{s} \approx 0$) is obtained by taking $\lambda \to \infty$, which yields an infinite duration but the smallest possible expended work $W_{\rm QS}={1 \over 2}D\gamma\ln(\kappa_f /\kappa_i)$, equal to the free energy difference between the two equilibria, as expected for a quasi-static process.
For finite $\lambda$ and making use of Eq. \eqref{eq:kappa-s}, Eq. \eqref{eq:variance} can be rewritten as $\dot{s}= -2\sqrt{s/\gamma\lambda}$, which possesses the general solution
\(
s(t) = (\sqrt{s_i}-t /\sqrt{\gamma\lambda})^2.
\)
Inserting this expression into Eq. \eqref{eq:kappa-s} yields the optimal evolution of the trap stiffness
\be
\kappa(t) = \frac{D\gamma + \sqrt{\gamma s_i/\lambda}-t /\lambda}
{\left(\sqrt{s_i}-t /\sqrt{\gamma\lambda}\right)^2}\,,
\label{eq:kappa}
\ee
which defines our protocol for the optimized transfer.

It is important to stress that, in our case, the Euler-Lagrange equation \eqref{eq:EL} is purely algebraic, so that one cannot enforce the initial and final conditions. Thus, except for the quasi-static limit, Eq. \eqref{eq:kappa} does not satisfy the conditions for which the system is at thermal equilibrium in the initial and final states. In order to ensure that $s_{i,f}\kappa_{i,f}=D\gamma$, we need to add to Eq. \eqref{eq:kappa} two discontinuities (as already noticed in \cite{SchmiedlPRL2007,SchmiedlEPL2008}). The optimal protocol thus consists of three successive sequences:
\begin{enumerate}
\item
At time $t=t_i$, the trap stiffness is suddenly changed from $\kappa_i = D\gamma/s_i$ (initial equilibrium) to $\kappa(t_i^{+}) = \kappa_i^{+}$, such that:
$\kappa_i^{+}- \kappa_i = \sqrt{\gamma/(\lambda s_i)}$, while keeping the variance equal to $s_i$;
\item
Between $t_i^{+}$ and $t_f^{-}$, the stiffness varies according to Eq. \eqref{eq:kappa}, reaching
\(
\kappa(t_f^{-})\equiv\kappa_f^{-} = D\gamma/s_f +\sqrt{\gamma /(\lambda s_f)}
\);
\item
At time $t=t_f$, the stiffness is suddenly changed from $\kappa_f^{-}$ to $\kappa_f = D\gamma/s_f$ (final equilibrium), while keeping the variance equal to $s_f$.
\end{enumerate}

\begin{figure}[htb]
  \centering{
    \includegraphics[width=1.0\linewidth]{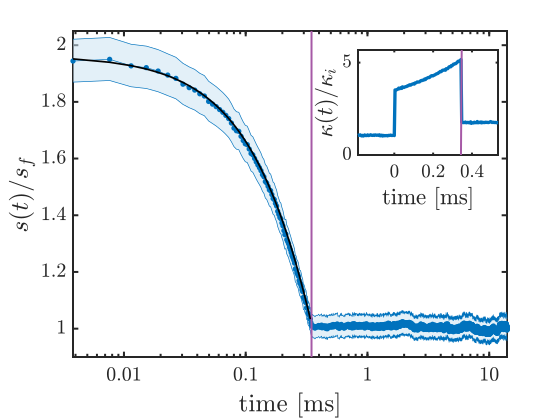}}
  \caption{Time-evolution of the variance $s(t)$ for an optimal protocol of duration of $\Delta t  = 3.47\times 10^{-4} \ {\rm s}\sim \tau_{\rm relax}/10$, indicated by the vertical solid line. Starting at time $t_i$ at thermal equilibrium with $\kappa_i = 2.77 \pm 0.08 \ {\rm pN}/\mu{\rm m}$ and a mean $s_i=1.48\times 10^{-15} \ {\rm m}^2$, the variance is extracted from the PDF and normalized to the final equilibrium state associated with the plateau-averaged value $s_f= 7.75\times 10^{-16} \ {\rm m}^2$ reached after $t_f$, corresponding to $\kappa_f = 5.22\pm 0.15 \ {\rm pN}/\mu{\rm m}$. The Lagrange multiplier $\lambda = (1.27\pm 0.02)\times 10^{17} \ {\rm s}/{\rm J}$ associated with this protocol is determined by the set of values $(\Delta t, s_i, s_f)$ and Eq. \eqref{eq:totaltime}. The superimposed black continuous line is the theoretical time-evolution of the optimized variance $s(t)=(\sqrt{s_i}-t/\sqrt{\gamma\lambda})^2$ calculated with the measured values. The experimental error bars correspond to a $95\%$ confidence interval for $s(t)$, including calibration uncertainties (see more details in Appendix A). The experimental optimal protocol $\kappa(t)$, normalized to the initial stiffness $\kappa_i$, is displayed in the inset.}
  \label{fig:1}
\end{figure}

Experimentally, we have performed $N=2\times 10^4$ successive and identical optimal protocols on the trapped microsphere, each built on the above three sequences, forcing the system to relax to thermal equilibrium within a time $\Delta t$ chosen to be shorter than $\tau_{\rm relax}$. The time-evolution of $s$ between two thermal equilibrium configurations is displayed in Fig. \ref{fig:1} for a shortened duration $\Delta t \sim\tau_{\rm relax}/10$, where $\tau_{\rm relax}=2 \gamma/\kappa_f= 3.22 \pm 0.09 \ {\rm ms}$. The reduction in $s(t)$ from its initial value corresponds to the fact that the trap is stiffer at the end of the protocol with $\kappa_f / \kappa_i \sim 1.85$. The time evolution of $s$ calculated using these values for $s_i$ and $\lambda$ is in excellent agreement with the experiment.

\begin{figure}[htb]
  \centering{
    \includegraphics[width=1.0\linewidth]{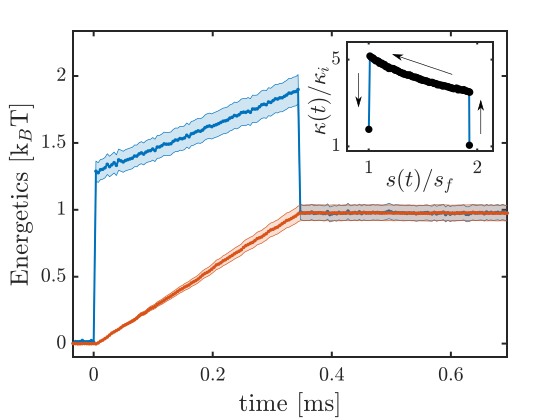}}
  \caption{Energetics associated with the optimal protocol described in Fig. \ref{fig:1} for $\Delta t \sim \tau_{\rm relax}/10$. The ensemble-averaged cumulative work $W(t)$ and heat $Q(t)$ are measured with respect to the initial thermal equilibrium. The experimental error bars are determined by the propagation of variance and calibration (i.e., stiffness) uncertainties (see the Appendix A). At $t>t_f$, the system has reached a thermal steady state with $Q  \sim  W $ -- see maint text. The control parameter $\hat{\kappa}(s)$ is displayed in the inset, the arrows corresponding to the time evolution.}
  \label{fig:2}
\end{figure}

This reduction in the transfer duration has an energetic cost that can be evaluated for each sequence using Eq. \eqref{eq:work}. Such cost is measured experimentally by evaluating the cumulative work $W(t)$ from the recorded evolution of $\hat{\kappa}(s)$. As seen in Fig. \ref{fig:2}, the time-evolution of $W(t)$ can also be split into three sequences. First,
the  quantity of work $W^{(1)}=s_i(\kappa_{i}^{+}-\kappa_i)/2$ is injected instantaneously into the system at the time $t_i$ as the trap is suddenly stiffened from $\kappa_{i}$ to $\kappa_{i}^{+}$.
During the second sequence, the injection of work continues as the trapping volume is progressively reduced, reaching $W^{(2)}=W_{\rm QS}+\sqrt{\gamma/\lambda}\left[(s_i^{3/2}-s_f^{3/2})/3-(s_i^{1/2}-s_f^{1/2})/2\right]$ at $t=t_f^{-}$, when $\kappa = \kappa_{f}^{-}$.
Finally, the trap is suddenly expanded at $t=t_f$, and the system instantly reaches its final equilibrium state, delivering to the thermal bath a quantity of work equal to $W^{(3)}=s_i(\kappa_f-\kappa_{f}^{-})/2$. For $t>t_f$, the thermal steady state is characterized by $W=Q$, with $W(t>t_f)=(0.981\pm 0.059) \ k_{\rm B}T$ and $Q(t>t_f)=(0.983\pm 0.060) \ k_{\rm B}T$.

In contrast, heat is continuously dissipated from the microsphere to the thermal bath, as seen in Fig. \ref{fig:2} from the monotonic increase of the ensemble-averaged cumulative heat $Q(t)$ throughout the protocol. The evolution of the dissipated heat between the two equilibrium states is almost exactly linear in time, which corresponds to a constant production of entropy.
We stress again that the experimentally measured values of the heat and work injected in and extracted from the system are in excellent agreement with the theoretical predictions.

The total work expended throughout the optimal protocol is evaluated by adding the contributions from each of the three steps described above. One obtains:
\be
W_{\rm opt} = W_{\rm QS}+ \sqrt{\gamma / \lambda}\,\left(s_i^{1/2}-s_f^{1/2}\right).
\label{eq:totalwork}
\ee
Similarly, the total duration of the optimal protocol is obtained by inserting Eq. \eqref{eq:kappa-s} into Eq. \eqref{eq:deltat}, yielding:
\be
\Delta t_{\rm opt} = \sqrt{\gamma\lambda}\, \left(s_i^{1/2}-s_f^{1/2}\right)
\, .
\label{eq:totaltime}
\ee
The above expressions clearly show that our optimization procedure is perfectly symmetric as far as duration and work are concerned and that the trade-off between these two quantities is governed by the Lagrange multiplier $\lambda$. Indeed, one can choose $\lambda$ using Eq. \eqref{eq:totaltime} to fix the total duration and then the minimum expended work will be given by Eq. \eqref{eq:totalwork}; or, alternatively, one can determine $\lambda$ through Eq. \eqref{eq:totalwork} to fix the total work and then the minimum duration of the process will be given by Eq. \eqref{eq:totaltime}.

This leads us to define the ``excess work'' of the optimal protocol as $\Delta W_{\rm opt} \equiv W_{\rm opt}-W_{\rm QS}$ and to note that the product
\be
\Delta t_{\rm opt}\,\Delta W_{\rm opt} = {\gamma \over 2} \,\left(s_i^{1/2}-s_f^{1/2}\right)^2
\label{eq:uncertainty}
\ee
is independent of $\lambda$ and only depends on the initial and final states. This equality fixes the mutually exclusive relation between transfer duration and expended work under optimal control. It corresponds to the frontier value of a universal exclusion region $\Delta t\,\Delta W \ge \gamma / 2(\sqrt{s_i}-\sqrt{s_f})^2$ that bounds from below all protocols that are not optimal.

\begin{figure}[htb]
  \centering{
    \includegraphics[width=1.0\linewidth]{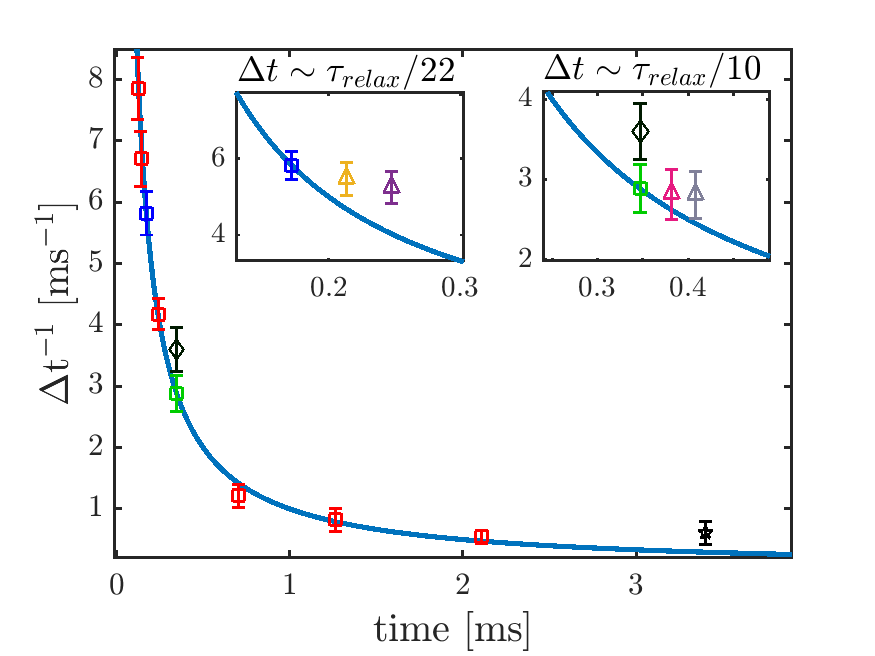}}
  \caption{Extracted excess works $\Delta W_{\rm opt}$ for a series (square) of optimal protocols defined by transfer durations $\Delta t=\tau_{\rm relax}/n$, with successively $n \sim 34, 30, 22, 16, 10,6, 3$ and 2, revealing the mutually exclusive relation between $\Delta W_{\rm opt}$ and $\Delta t$. For each $n-$ protocol, we normalize $\Delta W_{\rm opt}$ by the corresponding value $\gamma/2 (\sqrt{s_i}-\sqrt{s_f})^2$, considering that the precise values for $(s_i,s_f)$ slightly vary from protocol to protocol. The universality of the bound is clearly verified experimentally by observing that all optimized coordinates $\{\Delta t,\Delta W_{\rm opt}/(\gamma/2 (\sqrt{s_i}-\sqrt{s_f})^2)\}$ precisely fall (within error bars) on the $1/\Delta t$ curve. The excess work measured for an ``engineered swift equilibration" protocol \cite{MartinezNatPhys2016} (black diamond) defined for $\Delta t  = 3.47\times 10^{-4} \ {\rm s}\sim \tau_{\rm relax}/10$, and the excess work measured for a step-like protocol (black star) at $\tau_{\rm relax}\sim 3.22$ ms clearly fall above the optimal bound -- see Appendix B for details. Insets: $(\Delta t,\Delta W)$ coordinates measured for smooth (thus suboptimal) protocols for  $n\sim 22$ and $n\sim 10$, and smoothness parameters $\varepsilon =5\times 10^{-6}$, $\varepsilon =10^{-6}$, and $\varepsilon =0$, expressed in units of $ s_i ^2/(D\kappa_i^2)$. Such smooth protocols are defined using the same Lagrange multiplier $\lambda$ as their associated optimal protocols. For each case, the product $\Delta t\Delta W$ converges to the optimal lower bound (solid blue line) as $\varepsilon \to 0$. For $n\sim 10$, the excess work for the ESE protocol is plotted again (black diamond).}
  \label{fig:3}
\end{figure}

The frontier can be drawn experimentally by changing the transfer duration $\Delta t$ within the conditions of optimal control, i.e. changing the Lagrange multiplier.
To do so, we have measured $\Delta W_{\rm opt}$ for a series of eight optimal protocols with different $\Delta t$. By normalizing each measured value of $\Delta W_{\rm opt}$ to the associated value of $\gamma / 2(\sqrt{s_i}-\sqrt{s_f})^2$, one can test the universal nature of the bound. This is clearly confirmed in Fig. \ref{fig:3}, with all the optimal solutions implemented experimentally falling precisely on the $1/\Delta t$ curve. To further prove that the frontier corresponds to a lower-bound, we have verified experimentally that the ($\Delta t ,\Delta W)$ coordinates of typical non-optimal protocols -- continuous (see below), step-like, and ``engineered swift equilibration" protocols (see Appendix B) -- all fall above the expected bound, as displayed in Fig. \ref{fig:3}.

A salient feature of our optimal control procedure is represented by the sudden jumps in stiffness that have to augment the solution of Eq. \eqref{eq:kappa} in order to comply with thermally equilibrated initial and final configurations. From an experimental point of view, such discontinuities do not constitute a weakness of the procedure, as they correspond to finite and measurable quantities of work exchanged between the bath and the system \cite{AurellPRE2012,PlataPRE2019}. But it is interesting to stress that one asset of our variational strategy is its capacity to construct smooth protocols that are as close as desired to the optimal ones. For this, we need to control the derivatives of the function $\hat{\kappa}(s)$, which can be done by adding the gradient term $\int_{s_i}^{s_f}|\hat{\kappa}'(s)|^2 ds$ to the functional $J[\hat{\kappa}(s)]$ in Eq. \eqref{eq:J}, with a second Lagrange multiplier $\varepsilon$. Hence, we arrive at the modified Euler-Lagrange equation:
\be
2 \varepsilon  \, \frac{d^2 \hat{\kappa}}{ds^2} = \frac{\gamma s}{(D\gamma - s \hat{\kappa})^2}  - \lambda ,
\label{eq:ele}
\ee
which can be solved numerically as a boundary value problem, with initial and final conditions at thermal equilibrium $\hat{\kappa}(s_{i,f})=D\gamma/s_{i,f}$. Once the solution $\hat{\kappa}(s)$ is known, the time-evolution of the variance $s(t)$ is found by integrating Eq. \eqref{eq:variance} (more details are given in Appendix C).

\begin{figure}[htb]
  \centering{
    \includegraphics[width=1.0\linewidth]{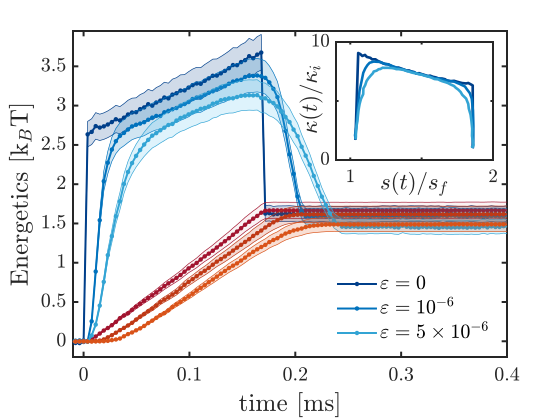}}
  \caption{Comparison of the cumulative energetics [expended work $W(t)$ (blue curves) and dissipated heat $Q(t)$ (red curves)] between an optimal protocol and two smooth protocols with $\varepsilon=5\times 10^{-6}$ and $\varepsilon=10^{-6}$, expressed in units of $ s_i ^2/(D\kappa_i^2)$, and identical value of $\lambda=(2.97\pm0.12)\times 10^{16} \ {\rm s}/{\rm J}$. As seen on the insets of Fig. \ref{fig:3}, although the smooth protocols involve slightly less work than the optimal one, they correspond to longer transfer durations. Inset: superimposed evolutions $s(t)$ vs. $\kappa(t)$ for the three protocols, showing the continuous nature of the smooth protocol and illustrating the progressive convergence to the optimal protocol in the $\varepsilon\rightarrow 0$ limit. For each protocol, the curves are  normalized to the corresponding $\kappa_i$ for $\kappa(t)$ and $s_f$ for $s(t)$.}
  \label{fig:4}
\end{figure}

 Using the same values of $\lambda$ that defined the optimal protocols with, respectively, $\Delta t \sim \tau_{\rm relax}/22$ and $\Delta t \sim \tau_{\rm relax}/10$ (Fig. \ref{fig:3}, inset), we implemented two smooth protocols for two different values of the Lagrange multiplier $\varepsilon=5\times 10^{-6}$ and $\varepsilon=10^{-6}$ (here and in the following,  $\varepsilon$ is expressed in units of $ s_i ^2/(D\kappa_i^2)$). As shown in Fig. \ref{fig:4}, the smooth protocols follow closely the optimal ones, except near the beginning and the end of the process, where they approach the equilibrium states in a continuous way. For the same value of $\lambda$,  smooth protocols give slightly longer transfer durations ($2.48\times 10^{-4} \ {\rm s}$ for $\varepsilon = 5\times 10^{-6}$ and $2.14\times 10^{-4} \ {\rm s}$ for $\varepsilon = 10^{-6}$) than the optimal protocol  ($\Delta t=1.72\times 10^{-4} \ {\rm s}$) but, as expected, the expended work is slightly smaller ($1.36\pm 0.06~k_{\rm B}T$ for $\varepsilon = 5\times 10^{-6}$ and $1.65\pm 0.06~k_{\rm B}T$ for $\varepsilon = 10^{-6}$) in the smooth case than in the optimized limit ($1.69\pm 0.06~k_{\rm B}T$). The non-optimal character of the smooth protocols is clearly seen in the insets of Fig. \ref{fig:3}, where all ($\Delta t ,\Delta W)$ coordinates lie above the universal bound, and only converge towards it in the $\varepsilon\rightarrow 0$ limit.

In conclusion, we have devised a family of optimal protocols that transfer an optically trapped microsphere between two equilibrium positions, minimizing both the transfer duration and the associated energetic cost.
Within such protocols, the trade-off between duration and work can be modulated at will by tuning a single Lagrange multiplier given by our variational approach.
A key result of our work is to show that the product $\Delta t\, \Delta W$ is bounded from below, in a way reminiscent of energy-time uncertainty relations.
Similar bounds were noticed in earlier works \cite{SekimotoBook,CilibertoPRX2017}, but only for some special cases. Here, our bound is universal (it depends exclusively on the initial and final states) and is only reached for the optimal protocol, as we demonstrated both theoretically and experimentally.
Further extending the present results to quantum systems may open new interesting perspectives in the burgeoning field of quantum stochastic thermodynamics
\cite{Elouard2017,vinjanampathyContPhys2016,RossnagelScience2016,CavinaPRA2018}.

\section{Acknowledgments}

This work was supported in part by Agence Nationale de la Recherche (ANR), France, ANR Equipex Union (Grant No. ANR-10-EQPX-52-01), the Labex NIE projects (Grant No. ANR-11-LABX-0058-NIE), and USIAS within the Investissements d'Avenir program (Grant No. ANR-10-IDEX-0002-02). Y. R.-C. is a member of the International Doctoral Program of the Initiative d'Excellence of the University of Strasbourg, whose support is acknowledged. L. M. is supported by the National Natural Science Foundation of China, Research Fund for International Young Scientists under the
project No. 1161101053 and the Young Scientist Program under the project No. 11601335.

\section{Appendix A: Setup, calibration, uncertainties}

\subsection{Optical trap setup}

All experiments are performed on single optically trapped polystyrene spheres (radius $R=500$ nm) taken from a monodisperse ($\delta R /R = 0.028$) solution (ThermoFisher, FluoSpheres) and enclosed inside a fluidic cell filled with dionized water. The microfluidic cell is made with a microscope slide and a 170 $\mu$m thick glass coverslip, sealed with a 120 $\mu$m thick spacer.

\begin{figure}[htb]
  \centering{
    \includegraphics[width=0.9\linewidth]{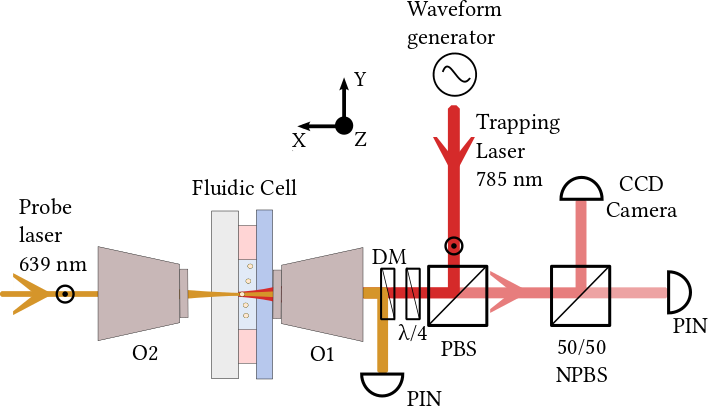}}
  \caption{The trapping laser ($\lambda_T=785$ nm, 100 mW, TEM$_{00}$, CW, Coherent, OBIS LX785) is modulated externally using a waveform generator (Agilent, 33220A). Linearly polarized along the $z-$axis, the beam is sent to a water-immersion objective (O1, $100\times$, 1.2 numerical aperture (NA)) through a polarizing beam splitter (PBS) and a quarter-wave plate ($\lambda/4$). The  intensity $ I(t)$ partially reflected by the end-surface of the fluidic cell varies linearly with the displacement $x(t)$ of the polystyrene microsphere inside the trap. This intensity $I(t)$ is collected and recorded by a \textit{p-i-n} photodiode (Thorlabs, DET10A), while a CCD camera is used in the other port of the non polarizing beam splitter (NPBS) for imaging. The probe beam consists of a second laser ($639$ nm, $70$ mW Thorlabs laser diode, linearly polarized) of low power ($400~\mu$W). It is injected inside the trap collinearly with the trapping beam but from behind the fluidic cell using a dry objective (O2, $60\times$, NA 0.7). This second beam is separated from the trapping beam using a dichroic mirror (DM) and the interference between the transmitted beam and the diffracted light by the bead is recorded using a second \textit{p-i-n} photodiode (Thorlabs, DET10A) placed in a plane conjugated to the back focal plane of the trapping objective. In order to ensure that a single bead is trapped without other beads in its vicinity, potentially perturbing the dynamics, the optical trap is equipped with an interferometric scattering microscope not shown here but described in details in our previous work \cite{SchnoeringPRL2018}.}
  \label{fig:setup}
\end{figure}

The optical trap, described in details in Fig. \ref{fig:setup}, is an evolution of the setup described in our previous work \cite{SchnoeringPRE2015,SchnoeringPRAppl2019,SchnoeringPRL2018}. It uses a CW near-infrared ($\lambda_{T} = $785 nm) laser whose intensity -- hence the trap stiffness -- can be modulated externally using a waveform generator. Any trapping protocol can then be implemented by computer-programming the waveform generator so that the time-evolution of the trap stiffness follows the desired profile.

Under such trapping laser modulation, the instantaneous axial motion $x(t)$ of the bead is monitored using an auxiliary laser propagating in the opposite direction of the trapping laser (see Fig. \ref{fig:setup}). We checked that this low-power probe beam, injected in the fluidic cell from its back-side, does not exerts any spurious optical force of the trapped bead. The signal collected by the photodiode and the output voltage of the waveform generator are simultaneously registered by a multichannel acquisition card (National Instruments, NI-6251) with a sampling rate $f_s = $ 2$^{18}$ Hz. In order to span the signal in the full dynamic range of the acquisition card, the generator output voltage was re-scaled using a scaling amplifier (Stanford Research Systems, SIM983) and the voltage time series of the photodiode was amplified and filtered using low-noise pre-amplifiers (Stanford Research Systems, SR560).

\subsection{Stiffness modulation calibration}

\begin{figure}[htb]
  \centering{
    \includegraphics[width=1.0\linewidth]{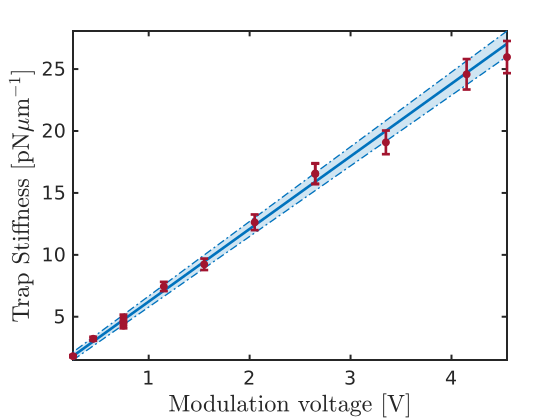}}
  \caption{Evolution of the trap stiffnesses as a function of DC waveform generator voltages. The red dots represent the stiffness values extracted from the motional PSD, with error bars for each point combining the uncertainties of the Lorentzian fit of each PSD and the error made on the Stokes drag $\gamma = 6\pi R \eta$ due to the polystyrene sphere radius dispersion $\delta R /R$. The solid line is the linear fit and the shaded area represents a 95 \% confidence interval for the estimated linear regression parameters taking into account the weights of the data points.}
  \label{fig:calibration}
\end{figure}

The trapping laser is modulated according to a given protocol $\kappa(t)$, defined and calculated with chosen transition parameters ($\ki,\kf, \Delta t$). In order to convert this protocol $\kappa(t)$ into a modulating voltage $V_{\rm mod}(t)$ for the waveform generator, a calibration procedure is performed. This procedure consists in measuring the trap stiffnesses associated with a series of consecutive values of DC voltages, i.e. consecutive trapping laser intensities. Each stiffness is extracted from a Lorentzian fit of the corresponding motional power spectral density (PSD) of the trapped bead. Associated error bars are obtained from the uncertainties of the Lorentzian fits (MATLAB Levenberg-Marquardt algorithm). The calibration curve shown in Fig. \ref{fig:calibration} corresponds to a linear fit of the evolution of such measured stiffnesses (including their error bars) as a function of the DC voltages.

\subsection{Monitoring Brownian dynamics}

\begin{figure}[htb]
  \centering{
    \includegraphics[width=1.0\linewidth]{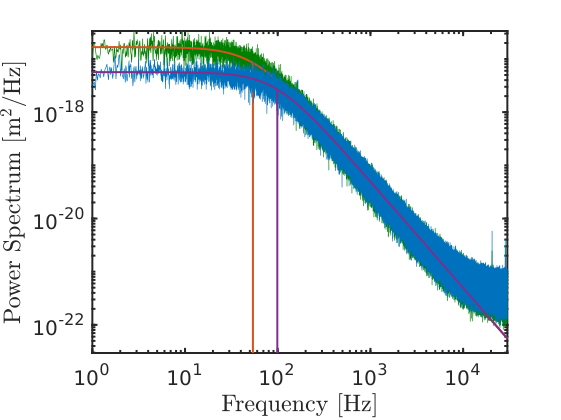}}
  \caption{The power spectral density of the concatenated trajectories corresponding to the sections of the cycles for which $\kappa$ is fixed to $\ki$ is displayed in greeen. The best-fitted roll-off frequency $f_c = 52.63 \pm 0.01 $ Hz (vertical red line) yields $\ki = 2.78 \pm 0.08$ \pN, and the position sensitivity parameter is $\beta = \sqrt{k_BT/\gamma D_{fit}} = 1.21 \pm 0.02 $ $\mu$m/V -see below. The blue curve is the power spectral density of the concatenated trajectories corresponding to the sections of the cycles for which $\kappa$ is fixed to $\kf$. The best-fitted roll-off frequency is $f_c = 98.98 \pm 0.02 $ Hz  (vertical purple line) gives $\kf = 5.22 \pm 0.15 $ \pN for this case. Here, the positional calibration factor is $\beta = 1.31 \pm 0.02 $ $\mu$m/V. Lorentzian fits (continuous red and purple lines superimposed to the PSDs) are calculated by implementing a MATLAB Levenberg-Marquardt algorithm for non-linear leasts squares.}
  \label{fig:ki_psd}
\end{figure}

The time evolution of the Brownian system is monitored by recording the stochastic trajectory of the trapped bead over $2 \times 10^4$ cycles of the protocol $\kappa(t)$. Each cycle lasts 50 ms, where the first 30 ms correspond to the initial thermal equilibrium with $\ki$ and the remaining ($20 - \Delta t$) ms correspond to the final thermal equilibrium at $\kf$.  Each stationary region of the full trajectory, i.e. corresponding to a constant $\kappa$  ($\ki$ or $\kf$ ), is sectioned and concatenated with all the other sliced trajectories under the same stiffness.  The PSD of this concatenated trajectory is computed and a Lorentzian fit yields the ensemble average $\kappa$. Figs. \ref{fig:ki_psd} (a) and (b) respectively show the PSD of the concatenated trajectories for the equilibria $\ki$ and $\kf$ for the case $\Delta t \sim \tau_{\rm relax}/10$ described in the main text.

Implementing the same procedure, the full temporal trace of the particle positions undergoing $2 \times 10^4$ cycles is chopped into trajectories that correspond to a single cycle of the protocol $\kappa(t)$. The ensemble of traces then consists of all the sub-trajectories superimposed within the same time interval, in such a way that they all start $t = -30$ ms with $\ki$, as displayed in Fig. \ref{fig:trajectories} below.

\begin{figure}[htb]
  \centering{
    \includegraphics[width=1.0\linewidth]{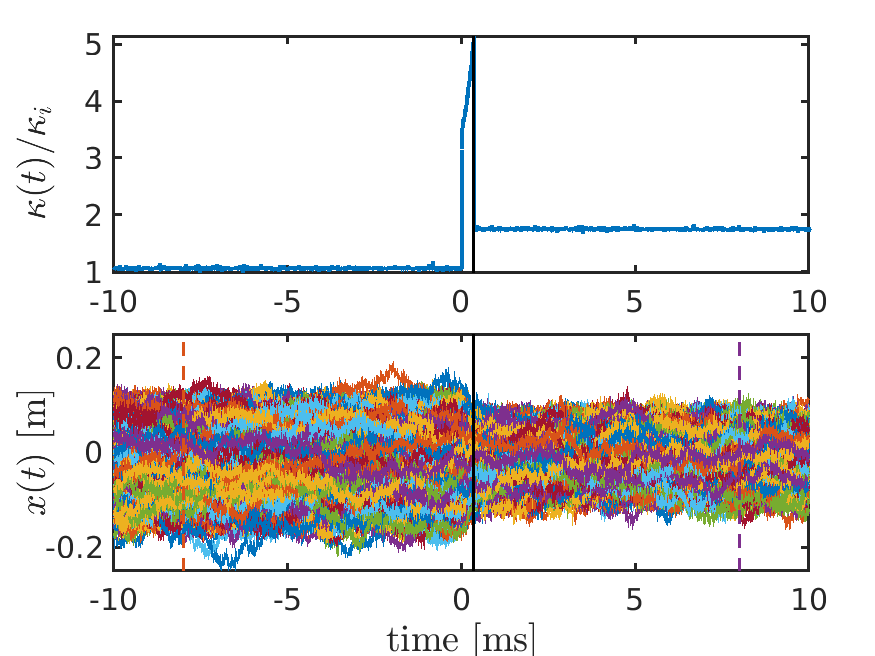}}
  \caption{ Ensemble of trajectories corresponding to one cycle. Top: A single cycle of the control parameter $\kappa(t)$ normalized to $\ki$. Bottom: Position fluctuations of the bead in the trap of modulated stiffness. The solid vertical lines indicate $\Delta t = 3.47 \times 10^{-4}$s. The position distribution functions calculated at the two times indicated by the dashed vertical lines in the lower panel are displayed in Fig. \ref{fig:gauss} below (top panel).}
  \label{fig:trajectories}
\end{figure}

\begin{figure}[htb]
  \centering{
    \includegraphics[width=1.0\linewidth]{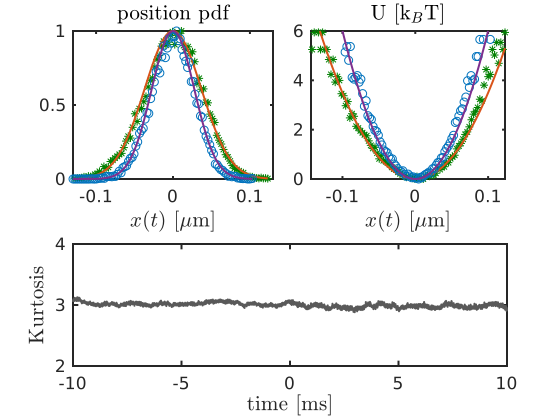}}
  \caption{Top-left panel:  Position distribution functions (PDF) built from the ensemble of trajectories at the two different times $t_j < t_0$  and $t_j > t_f$ indicated by the two dashed vertical lines in Fig. \ref{fig:trajectories} above (bottom panel), with associated trap stiffnesses $\ki$ and $\kf$ respectively. Top-right panel:  Associated trapping potentials extracted from the PDF as $U(x,t_j) = - k_{\rm B} T \log (\rho(x,t_j)) + {\rm cst}$. The solid lines correspond to $U = \frac{1}{2}\kappa x^2$ with $\kappa=\kappa_i$ and $\kappa=\kappa_f$ extracted from the PSD shown in Fig. \ref{fig:ki_psd}. Bottom panel: Kurtosis of each PDF for all times $t_j$.}
  \label{fig:gauss}
\end{figure}

The instantaneous ensemble variance $s(t_j)$ at a time $t = t_j$ ($j = 1,\cdots , T\times f_s$), with $T=50$ ms and $f_s = 2^{18}$ Hz) is obtained by a vertical cross-cut of the ensemble of trajectories plotted in Fig. \ref{fig:trajectories}. The resulting distribution of positions $\rho(x,t_j)$ is a Gaussian of zero mean $\mu_x(t_j)$ and variance $s(t_j)$. Fig. \ref{fig:gauss} displays the position distribution functions (PDF) before (equilibrium at $\kappa_i$) and after (equilibrium at $\kappa_f$) the change in trapping stiffness imposed by the protocol $\kappa(t)$. The corresponding trapping potentials calculated as $U(x,t_j) = - k_{\rm B} T \log (\rho(x,t_j)) + {\rm cst}$ are also shown and compared to the expected harmonic profiles $U = \frac{1}{2} \kappa x^2$ evaluated from the stiffnesses $\kappa_i,\kappa_f$ that were extracted from the measured PSD shown in Fig \ref{fig:ki_psd}.

Proceeding in the same manner but for all times $t_j$, we can obtain the temporal evolution of the ensemble variance $s(t)$ over the full protocol $\kappa(t)$. To confirm that all PDF remain Gaussian for all times, we calculate their kurtosis and verify -see Fig. \ref{fig:gauss}, bottom panel- that all-time kurtosis remain very close to $3$ throughout the entire protocol.

\subsection{Statistical uncertainties}

The uncertainties for the instantaneous ensemble variances are obtained following a $\chi^2$ law with $N-1$ degrees of freedom where $N = N_{cycles}$ is the number of independent trajectories $x_i(t)$ undergoing one cycle of the protocol $\kappa(t)$.

\subsection{PSD calibration uncertainties}

Under a trapping laser intensity, the registered \textit{p-i-n} voltage values $V(t)$ that correspond to the position fluctuations of the trapped bead are converted into displacement units using the best-fit parameter of the Lorentzian fit of the PSD of the trajectory (at constant $\kappa$).  The fit parameter $D_{\rm fit}$ is compared to the diffusion coefficient $D = k_BT/\gamma$ expected from the Fluctuation-Dissipation Theorem, assuming known temperature and viscosity. This gives a conversion factor $\beta = \sqrt{D/D_{\rm fit}} $ from \textit{p-i-n} voltages to meters. The uncertainty on the position sensitivity is obtained  from standard error propagation including the uncertainty on the viscosity resulting from the $\delta R/R=2.8 \%$ size dispersion deviation of the trapped beads.

Instantaneous positions are thus given from the conversion factor as $x(t) = (\beta \pm \delta \beta) V(t)$, and therefore the variance, up to first-order in uncertainty, $x^2(t) = (\beta^2 \pm 2\beta\delta \beta)V^2(t)$, (since $\mu_x(t) = 0$). The total error of the variance writes as:
\begin{equation}
s(t_j) = \sigma^2_x(t_j) \pm \LaTeXunderbrace{(\delta \sigma^2_{\chi^2}(t_j) + \beta \delta \beta \sigma^2_x(t_j))}_{\delta s(t_j)},
\end{equation}
where $\sigma^2_x(t_j) = \sum^{N}_{i=1} \vert x_i(t_j) - \mu(t_j) \vert^2 / (N-1)$ is the estimator of the instantaneous ensemble variance over $N$ cycles, $\delta \sigma^2_{\chi^2}$ corresponds to the statistical uncertainty in the motional variance determination (see above) and $\delta \beta \sigma^2_x$ the PSD calibration uncertainty just discussed.

The temporal average variances related to the initial an final stiffness $s_i$ and $s_f$ are obtained from temporal average. Assuming $\Delta t$ as the interval over which $\kappa(t)$ remains constant (either at $\ki$ or $\kf$), the temporal average of the corresponding variance is:
\begin{equation}
\langle s \rangle_t = \frac{1}{\Delta t} \sum^n_{j = 1} s(t_j),
\end{equation}
taking $\Delta t$ as the interval over which $\kappa(t)$ remains constant (either at $\ki$ or $\kf$) and $n = \Delta t \cdot f_s$ with $f_s = 2^{18}$ Hz, the sampling frequency. The standard deviation of the temporal average is simply evaluated as:
\begin{equation}
\delta_t \langle s \rangle = \sqrt{ \frac{1}{\Delta t } \sum^n_{j = 1} \vert s(t_j) - \langle s\rangle_t \vert^2}
\end{equation}

The stationary variances $s_i$ and $s_f$ and their uncertainties are thus simply given by:
\begin{equation}
s_{i,f} = \langle s \rangle_t \pm  \LaTeXunderbrace{\left(\delta_t\langle s\rangle + \langle \delta s \rangle_t + \delta_t \langle \delta s\rangle\right)}_{\delta_t s_{i,f}},
\end{equation}
where $\langle \delta s \rangle_t =1/\Delta t \sum^n_{j = 1} \delta s(t_j)$.

\subsection{Energetics uncertainties}

The confidence interval of the mean cumulative work are computed taking into account the uncertainties related to both variances and stiffnesses. They are displayed on all energetic figures at a $95\%$ confidence level.

\section{Appendix B: Comparing optimal, step-like and ESE protocols}

We compare here three protocols that transfer the bead between two equilibria, going from an initial stiffness $\kappa_i$ to a final one $\kappa_f$ with, for all protocols, fixed and identical $\kappa_f , \kappa_i$ values given in the main text.

\begin{figure}[htb]
  \centering{
    \includegraphics[width=1.0\linewidth]{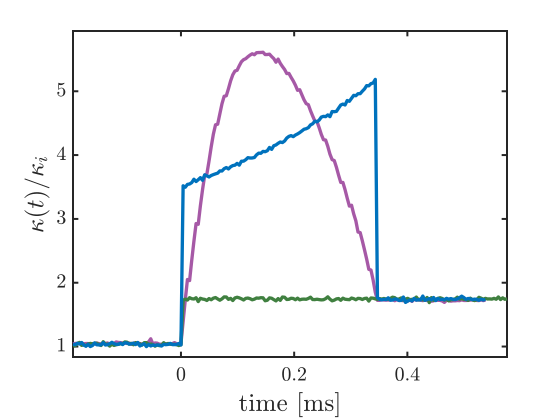}}
  \caption{Calibrated signal of the function generator, for a step-like (green), ESE  (pink), and optimal  (blue) protocols. The stiffness $\kappa(t)$ is normalized to the initial stiffness $\ki$. The jump for the transition $\ki \rightarrow \kf$ starts at $t_0 = 0$ s and, for the case of ESE and optimal ends at $\Delta t = 3.47 \times 10^{-4}$ s, with $\kappa_i = 2.77\pm 0.08, \kappa_f=5.22\pm 0.15$ pN$\mu$m. The ESE protocol $\kappa_{ESE}$ was computed based on Eq. (8) in \cite{MartinezNatPhys2016}. }
  \label{fig:protocols}
\end{figure}

The first protocol consists of a sudden step-like change of the optical trap stiffness -- see Fig. \ref{fig:protocols}, green trace. The second protocol is the ``engineered swift equilibriation'' (ESE) protocol recently proposed and implemented by Martinez, \textit{et al.}  \cite{MartinezNatPhys2016}. We calculate $\kappa_{\rm ESE}(t)$ following \cite{MartinezNatPhys2016} for a transfer duration of $\Delta t = 3.47\times 10^{-4}$ s. Over the same transfer duration, we also implement our optimal protocol $\kappa_{\rm opt}(t)$. All protocols are displayed in Fig. \ref{fig:protocols}.

\begin{figure}[htb]
  \centering{
    \includegraphics[width=1.0\linewidth]{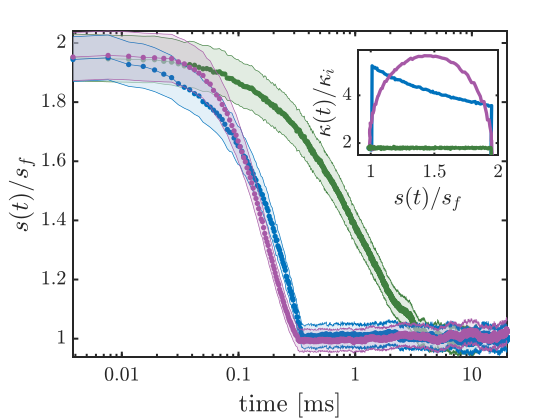}}
  \caption{Temporal evolution of the variance $s(t)$, after $t_0 = 0$ s for the step-like protocol (in green), and the ESE (in purple) and optimal (in blue) protocols. The variances are normalized to the final equilibrated variance $s_f$. The data points represent ensemble mean values of the variance $s(t)$ for each protocol. The shaded areas show the respective $95\%$ confidence intervals. Both ESE and optimal protocols reach an equilibrium regime $s_f$ at $\Delta t = 3.47 \times 10^{-4} \ {\rm s} \sim \tau_{\rm relax}/10 $ by construction. Inset: The control parameter $\hat{\kappa}(s)$ as a function of the variance $s$, with the same color codes as in the main figure.}
  \label{fig:vks}
\end{figure}

\begin{figure}[htb]
  \centering{
    \includegraphics[width=1.0\linewidth]{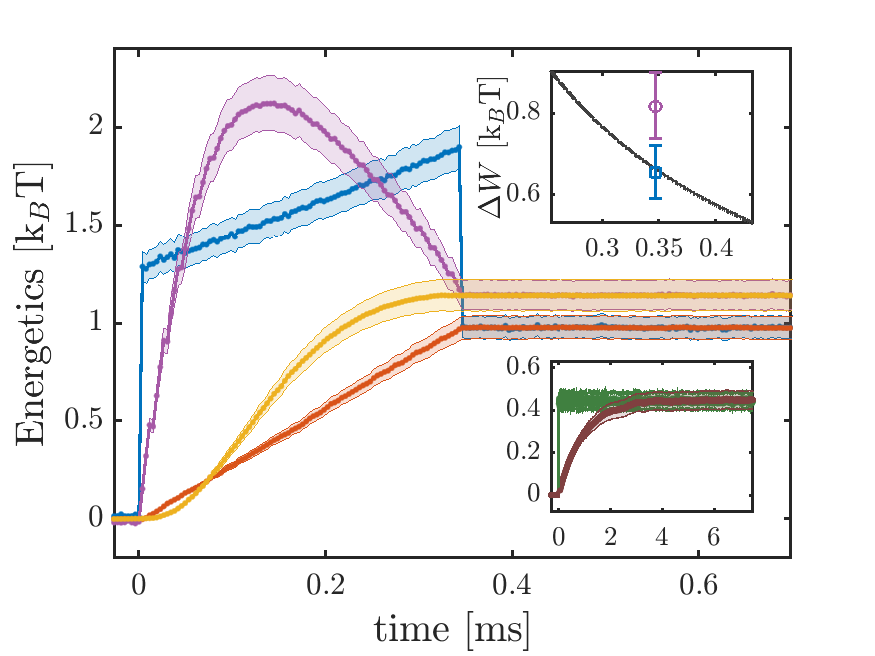}}
  \caption{Temporal evolution of the mean cumulative energetics of the different protocols, step-like (lower inset), ESE and optimal. The mean cumulative work for the optimal protocol is displayed in blue, with total work $W(t)_{\rm opt}= 0.981 \pm 0.059 \ k_{\rm B}T$. The mean cumulative work for the ESE protocol is displayed in pink, with total work $W(t)_{\rm ESE} = 1.142  \pm 0.075 \ k_{\rm B}T$. The mean cumulative heat generated through the optimal protocol is displayed in orange and the ESE protocol in yellow. Both are superimposed to the work, with total heat $Q(t)_{\rm opt} = 0.983 \pm 0.060 \ k_{\rm B}T$ and $Q(t)_{\rm ESE} = 1.142 \pm 0.076 \ k_{\rm B}T$. Shaded areas represent $95\%$ confidence levels. Lower inset: Energetics for the step-like protocol. As expected, the mean cumulative work (in green) reaches immediately  $W_{\rm step} = 0.45 \pm 0.04 \ k_{\rm B}T$. In brown, the heat, in contrast, achieves the equilibrium value $W  =  Q$ with $Q_{\rm step} = 0.45 \pm 0.04 \ k_{\rm B}T$  only after $\tau_{\rm relax}$. Upper inset: Comparison between the excess work values of the ESE protocol (pink) and the optimal one (blue) for the transfer duration of duration $\Delta t = t_f = 3.47 \times 10^{-4}$ s. The non-optimal character of the ESE protocol is directly measured with $\Delta W_{\rm ESE} = 0.81 \pm 0.08 \ k_{\rm B}T$ larger than the optimal value $\Delta W_{\rm opt} = 0.65 \pm 0.07 \ k_{\rm B}T$. The universal bound $\Delta W = \gamma (\sqrt{s_i}-\sqrt{s_f})^2/ \Delta t$ discussed in the main text is shown by the continuous line. }
  \label{fig:e2i}
\end{figure}

 Fig. \ref{fig:vks} gathers the time evolutions of the motional variances associated with each protocol. As expected, the step-like protocol displays the longest equilibration time when compared to the ESE and optimal protocols. From an energetic point of view, the comparison between the two latter protocols, shown in Fig. \ref{fig:e2i}, clearly reveals the non-optimal character of the ESE protocol with a cumulated work expense larger than for the the optimal protocol. This can also be seen in the inset of Fig. \ref{fig:e2i} where the excess work expended during the ESE protocol lies clearly above the optimal lower bound discussed in the main text.

\section{Appendix C: Smooth protocols}

The optimal protocol obtained in this work [Eq. (6) in the main text] was derived using the Lagrangian density
\be
L[s, \hat{\kappa}(s)] =\frac{\gamma }{D\gamma-s\, \hat{\kappa}(s)} - \lambda \hat{\kappa}(s)  .
\label{eq:C_Lagrange}
\ee
A peculiar feature of $L[s, \hat{\kappa}(s)] $  is that the corresponding Euler-Lagrange equation is purely algebraic (as opposed to a differential equation). Hence, it is not possible to impose the desired boundary conditions on the control parameter $\hat{\kappa}(s)$ (i.e. $s_i\kappa_i = s_f\kappa_f = D\gamma$) and two jumps have to be added ``by hand" at the beginning and the end of the protocol, as explained in the main text.

Although these jumps can be realized without much trouble in the experiments, it is interesting to develop a theoretical procedure capable of furnishing a suboptimal protocol $\hat \kappa(s)$ that is continuous in the variable $s$ and converges towards the optimal protocol as some parameter tends to zero.
To do this, we need to limit the gradient of $\hat{\kappa}(s)$ by adding a further term to the Lagrangian density \eqref{eq:C_Lagrange}, which becomes:
\be
L[s, \hat{\kappa}(s)] =\frac{\gamma }{D\gamma-s\, \hat{\kappa}(s)} - \lambda\hat{\kappa}(s) + \varepsilon |\hat{\kappa}'(s)|^2  \, ,
\label{eq:C_Lagrange1}
\ee
where $\varepsilon$ is an additional Lagrange multiplier. The above  Lagrangian density  yields the Euler-Lagrange equation (11) in the main text, which we reproduce here:
 \be
 2\varepsilon  \, \frac{d^2 \hat{\kappa}}{ds^2} =  \frac{\gamma s}{(D\gamma - s \hat{\kappa})^2} - \lambda .
\label{eq:C_ele}
\ee
As a second-order differential equation, Eq. \eqref{eq:C_ele} needs two independent boundary conditions, thus enabling us to set $s_i\kappa_i = s_f\kappa_f = D\gamma$, as requested for our protocols.
When $\varepsilon \to 0$, we obtain the correct limit case of Eq. (6) in the main text, i.e., the optimal protocol containing two points of infinite derivative (jumps) for the function $\hat{\kappa}(s)$ at $s_i$ and $s_f$.
 Through the Lagrange multiplier $\varepsilon$, one can limit  the value of such derivative, so that the protocol becomes smoother and smoother as $\varepsilon$  increases.

Equation \eqref{eq:C_ele} can be solved numerically by successive iterations. We used the following scheme:
\be
 -\alpha \hat{\kappa}_{i}^{n+1} + 2\varepsilon\left(\frac{d^2 \hat{\kappa}}{ds^2}\right)_i^{n+1} = \frac{\gamma s_i}{(D\gamma - s_i \hat{\kappa}_i^n)^2} - \lambda  -\alpha \hat{\kappa}_{i}^n ,
\label{eq:C_iterative}
\ee
where the superscript $n$ denotes the $n$-th iteration, while the subscript $i$ refers to the discrete grid $s_i = i\, \delta s$, with spacing equal to $ds$.
The second derivative is then approximated with the standard finite-difference formula:
\[
\left(\frac{d^2 \hat{\kappa}}{ds^2}\right)_i  \approx  \frac{\hat{\kappa}_{i-1} -2\hat{\kappa}_{i} + \hat{\kappa}_{i+1}  }{\delta s^2} .
\]
The parameter $\alpha>0$ is needed to ensure the convergence of the iterative procedure, but does not affect the final result (indeed it disappears from Eq. \eqref{eq:C_iterative} when $\hat{\kappa}_{i}^{n+1} =\hat{\kappa}_{i}^{n} $).

\begin{figure}[htb]
  \centering{
    \includegraphics[width=0.7\linewidth]{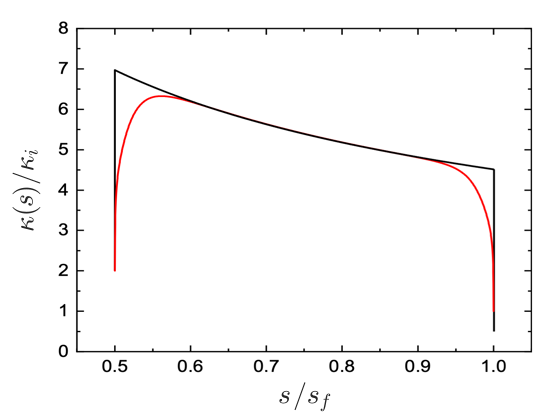}
    \includegraphics[width=0.7\linewidth]{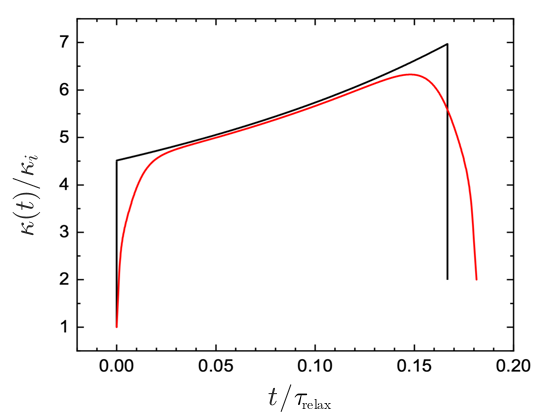}
    }
  \caption{Smooth protocol obtained from the solution of Eq. \eqref{eq:C_ele} (red lines) and corresponding optimal protocol with same value of $\lambda$ (black lines). Top panel: Protocols in the $(\hat{\kappa}, s)$ plane. Bottom panel: Protocols $\kappa(t)$ as a function of time. }
  \label{fig:appC}
\end{figure}

\begin{figure}[htb]
  \centering{
 \includegraphics[width=0.7\linewidth]{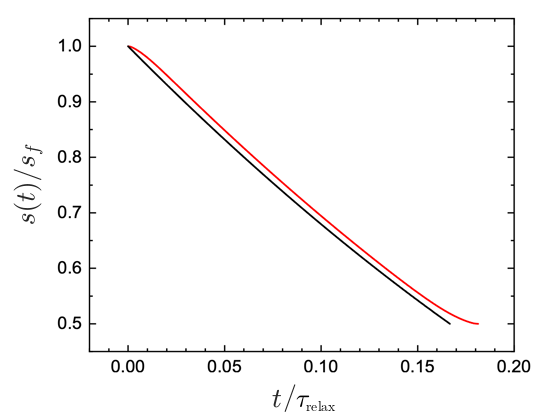}
 \includegraphics[width=0.7\linewidth]{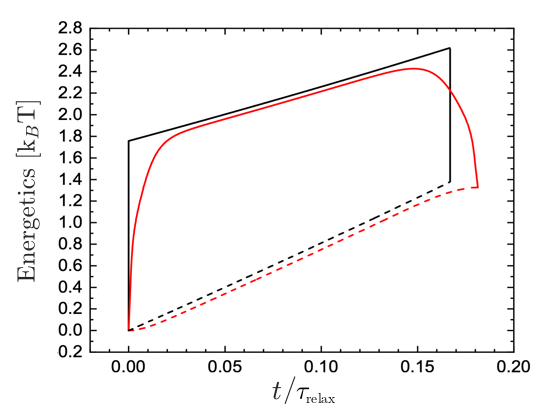}
    }
  \caption{Smooth protocol obtained from the solution of Eq. \eqref{eq:C_ele} (red lines) and corresponding optimal protocol with same value of $\lambda$ (black lines) . Top panel: Variance $s(t)$  as a function of time. Bottom panel: Dissipated heat $Q(t)  =-\frac{1}{2}\int_{t_i}^{t}{\rm d}t \dot{s}(t)\kappa(t)$ (dashed lines) and expended work $W(t) =\frac{1}{2}\int_{t_i}^{t}{\rm d}t s(t)\dot{\kappa}(t)$ (solid lines) as a function of time.}
  \label{fig:appC1}
\end{figure}

As an example, we have solved Eq. \eqref{eq:C_ele} with physical parameters $D=\gamma=1$ and $\lambda=0.81$, corresponding to a total duration for the optimal protocol $\Delta t_{\rm opt} \sim \tau_{\rm relax}/6$ according to Eq. (9) in the main text. The boundary values are $s_i=1$ and $s_f=0.5$, $\kappa_i=1$ and $\kappa_f=2$. The smoothness parameter is $\varepsilon =10^{-5}$.
The numerical convergence parameter is set to $\alpha=0.3$.
The result of the numerical integration is given in Figs. \ref{fig:appC} and \ref{fig:appC1},  for both the optimal (black lines) and smooth (red lines) protocols.
As expected, the smoothed protocol follows closely the optimal one, except near the extremities where it reaches its boundary values smoothly and without jumps.
The total time of the smoothed protocol is $0.182\times\tau_{\rm relax}$, longer than that of the optimal one. But the total work is smaller $W_{\rm smooth}=1.32< W_{\rm opt} = 1.38$. The time-energy product is $(\Delta t\, \Delta W)_{\rm smooth} = 0.356 >  (\Delta t\, \Delta W)_{\rm opt} = 0.343 $, in agreement with the theoretical considerations detailed in the main text.

\bibliography{biblio-optimal}

\end{document}